# Microscale characterisation of the time-dependent mechanical behaviour of brain white matter

Asad Jamal [a,*,#], Andrea Bernardini [a,#] and Daniele Dini [a]

[a] *Department of Mechanical Engineering, Imperial College London, London, SW7 2AZ, UK*

[*]Correspondence e-mail: a.jamal@imperial.ac.uk

[#]These authors contributed equally to the work presented in this article

## Abstract

Brain mechanics is a topic of deep interest because of the significant role of mechanical cues in both brain function and form. Specifically, capturing the heterogeneous and anisotropic behaviour of cerebral white matter (WM) is extremely challenging and yet the data on WM at a spatial resolution relevant to tissue components are sparse. To investigate the time-dependent mechanical behaviour of WM, and its dependence on local microstructural features when subjected to small deformations, we conducted atomic force microscopy (AFM) stress relaxation experiments on corpus callosum (CC), corona radiata (CR) and fornix (FO) of fresh ovine brain. Our experimental results show a dependency of the tissue mechanical response on axons orientation, with *e.g.* the stiffness of perpendicular and parallel samples is different in all three regions of WM whereas the relaxation behaviour is different for the CC and FO regions.

An inverse modelling approach was adopted to extract Prony series parameters of the tissue components, *i.e.* axons and extra cellular matrix with its accessory cells, from experimental data. Using a bottom-up approach, we developed analytical and FEA estimates that are in good agreement with our experimental results.

Our systematic characterisation of sheep brain WM using a combination of AFM experiments and micromechanical models provide a significant contribution for predicting localised time-dependent mechanics of brain tissue. This information can lead to more accurate computational simulations, therefore aiding the development of surgical robotic solutions for drug delivery and accurate tissue mimics, as well as the determination of criteria for tissue injury and predict brain development and disease progression.

# 1. Introduction

Brain is a biological system mainly composed of neurons and neuroglia and possess extreme complexity arising from understanding the functions that result from the interactions of about 86 billion neurons and their 100 trillion connections. Though brain studies are widely focused on biochemical or electrophysiological activity there is a clear evidence that classical mechanical concepts such as stress, deformation, pressure, strain and strain rate are crucial players in modulating both brain function and form (Goriely et al., 2015). Specifically, the time-dependent mechanical properties of cerebral white matter (WM) of brain tissue, a pathway for electric signals to travel from neuron to neuron through their axons, is important to understand to explore the underlying mechanisms and model brain deformations in traumatic brain injury, neurosurgery, introduction of implants or microcatheter insertion, tumour growth, brain development and targeted drug delivery (Cloots et al., 2013; Coppola et al., 2016; Pogoda and Janmey, 2018; Sharp et al., 2009).

The WM undergoing stretch and shear deformation is proposed to contribute to brain injury such as diffuse axonal injury (Feng et al., 2017a), impact other processes such as remyelination (Jagielska et al., 2012) in neurodegenerative or autoimmune diseases and respond to external stimuli *e.g.* implants due to mismatch of the mechanical compliance between cells and external devices with an inflammatory reaction and changes in cells morphology (Moshayedi et al., 2014).

Traditionally, studies on mechanics of WM are focused on the macroscale properties (Budday et al., 2017b, 2015; Cheng et al., 2008; Forte et al., 2017; Franceschini et al., 2006) and only a few attempts have been made to consider localised heterogeneous microstructural information and microscale properties (Feng et al., 2017a, 2013). However, recent studies have revealed localised microstructural differences such as axons volume fraction and geometry in specific regions *i.e.* corpus callosum (CC), corona radiata (CR) and fornix (FO) of brain WM (Bernardini et al., 2021). Furthermore, our recent work has shown that at relevant length scales the directionality of axons has significant effect on the localised properties *e.g.* drug infusion in corona radiata (Jamal et al., 2021) and permeability in different WM regions (Vidotto et al., 2019). This develops a need to investigate microscale mechanical properties of specific regions of WM to understand underlying localised mechanisms. There are attempts to investigate the microscale mechanics of WM however they either do not take in to account the anisotropy (Canovic et al., 2016) and lack detail investigations of regional differences in WM (Koser et al., 2015) or were performed on small animals such as mice (Koser et al., 2015; Moeendarbary et al., 2017; Sharp et al., 2009), not a closest resemblance to human brain. A summary of various findings in literature on the experimental mechanical characterisation of brain tissue is listed in Table S1, see Supplementary Information. To the best of our knowledge, a comprehensive analysis of animal brain WM with close resemblance to human brain, which considers the localised heterogeneities and directionality of WM, is still lacking in the literature.

Furthermore, developing realistic mechanical models (Forte et al., 2018; Terzano et al., 2021) for WM tissue, which account for its anisotropy (Feng et al., 2017b, 2013) and are based on accurately extracted material parameters and precise geometrical features, is crucial not only for studies like traumatic brain injury (Brands et al., 2004), but also for the development and successful implementation of robotic surgical tools such as micro catheters for convection-enhanced delivery systems (CED) (Dumpuri et al., 2007; Miller, 1999; Miller et al., 2000; Miller and Chinzei, 2002, 1997). Several studies have applied the theory of composites and a micromechanical approach to characterise biological materials (Arbogast and Margulies, 1999; Chen et al., 2013; Jorba et al., 2017). In central nervous system (CNS), this has been applied to certain regions, macroscopically characterized by a composite fibrous structure like the brainstem (Abolfathi et al., 2009; Karami et al., 2009; Yousefsani et al., 2018). However, no such efforts have been made to extract constituent material parameters of distinguish regions of brain WM such as CC, CR and FO from microscale experimental data and to develop mechanical model.

The current study aims to systematically characterise time-dependent micromechanics of CC, CR and FO using atomic force microscopy (AFM) and explore if axons directionality plays a role in mechanical behaviour of the WM tissue. The AFM is an established technique to investigate the mechanical properties of soft matter including cells and tissues (Darling et al., 2008, 2007; Garcia, 2020). The AFM requires only small volume of tissue and thus allow investigation of small regions such as FO, in contrast to classical rheological techniques, which require relatively large, and therefore inhomogeneous (in terms of axon bundle composition) samples. We focus on the characterisation of the linear elastic and viscoelastic response in configurations leading to small deformations, typically induced by localised pressures and pressure gradients found in CED procedures (Zhan et al., 2019), rather than developing a complete description of the nonlinear poroelastic constitutive behaviour required to capture the tissue response when subjected to *e.g.* large deformations and transient loading. Fully nonlinear poroelastic characterisation requires the use of relatively large indenter to probe coupled biphasic nonlinear response and will be the subject of a separate study. Nonetheless, this study aims to provide important benchmark data and a methodology that enables the characterisation of the different components of brain tissue in the small strain limit, which must also be matched by studies devoted to tissue characterisation under large deformations.

We choose sheep brain as the relatively large size and presence of sulci make gyrencephalic ovine brain functionally and anatomically more similar to human brain than lissencephalic brain of mice, rates and rabbits (FINNIE, 2001). The sheep brain exhibits resemblances to human brain in neurovascular structure (Hoffmann et al., 2014), in cytoarchitectural geometry (Bernardini et al., 2021), neuroradiological features (Boltze et al., 2008) and electroencephalographic records (Opdam et al., 2002), making it an excellent model in translational experimental neuroscience (Morosanu et al., 2019).

Furthermore, we analysed the AFM experimental data of CC, CR and FO as composite materials using a micromechanical approach, extracted Prony parameters of WM components *i.e.* axons and extra cellular matrix with its accessory cells forming the extra cellular matrix (ECM) when they are in their natural environment. This enables us to perform a FEA based micromechanics simulation on a representative volume element (RVE) of the tissue. This reveals how these individual components respond to external loading as a joint system and their contribution to the homogenised mechanical behaviour of WM without missing the effects of important interactions happening between these cytological components, and their important effect of the system response (Schregel et al., 2012; Weickenmeier et al., 2016; Young et al., 2016).

## 2. Materials and methods

### 2.1 Sample preparation

Fresh ovine brains were arranged from a local slaughterhouse and, using sharp surgical blades, slices of ~7-8 mm thickness were cut along two planes as shown in Fig. 1a. In such slices CC, CR and FO get exposed from two perpendicular directions (Pieri et al., 2019) as marked by hexagon, square and star respectively in Fig. 1b,c.

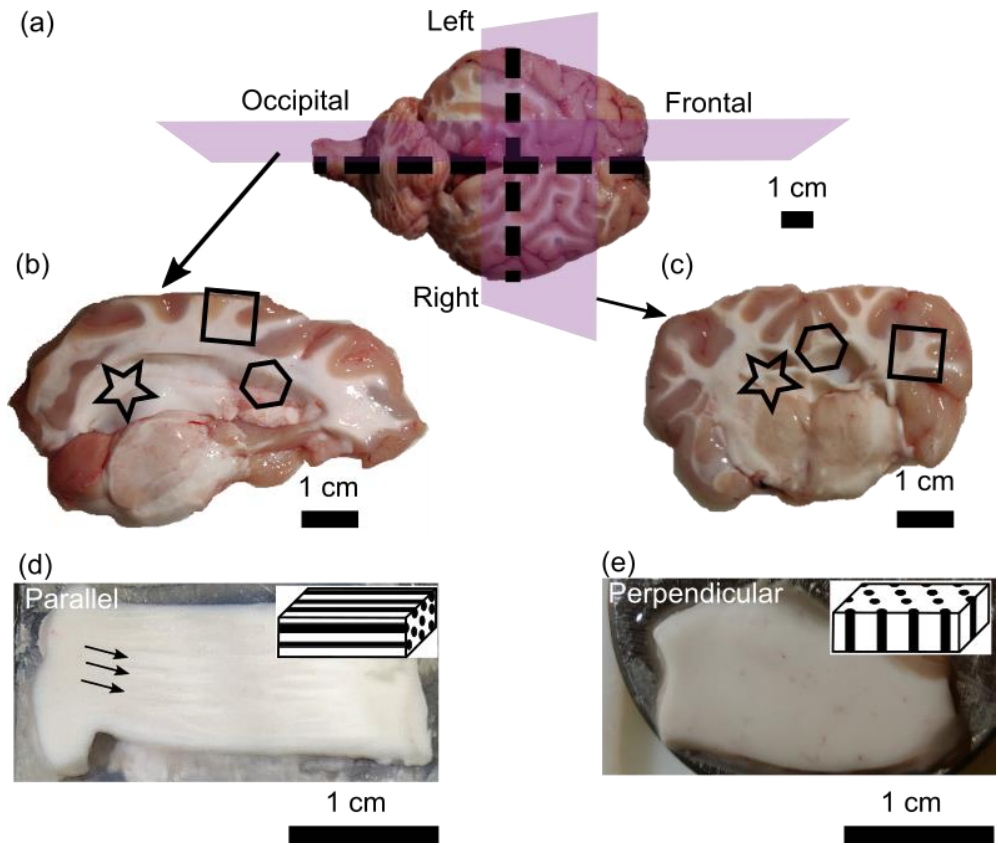

Fig. 1 (a) Fresh ovine brain, (b, c) slices cut along the two planes to expose different parts of WM in two perpendicular directions. Areas marked with hexagon, square and star are CC, CR and FO respectively. (d, e) Finely shaved samples where bundles of directional axons (shown in schematic) are visible in parallel sample, marked with arrows, and not visible in perpendicular sample.

This allowed sampling where axons were either parallel or perpendicular to the surface. During slicing, phosphate-buffered saline (PBS) was sprayed on the tissue to keep it hydrated. Finely shaved samples of ~ 2-3 mm thickness were made from CC, CR and FO using a vibratome. In such samples, axons were either parallel (parallel samples) or perpendicular (perpendicular samples) to the surface (Fig. 1d,e). Figure 1d shows a CC parallel sample where bundles of axons, marked with arrows, and inset is a schematic representation of parallel axons in matrix. Fig. 1e shows a CR perpendicular sample where no such directionality is macroscopically visible and inset is a schematic representation of perpendicular axons in the matrix. Please note, in case of CR, because of its macroscopic arrangement a more difficult visual inspection of the fibre directionality was encountered, and it was not always possible to have a perfect perpendicular sample. Finally, such samples were glued on plastic petri dish to avoid floating after being submerged in PBS, and microscale mechanical characterisation was performed using AFM.

## 2.2 Experimental protocol and data acquisition

AFM measurements were performed using a JPK Nanowizard CellHesion 200 AFM (JPK Instruments AG, Berlin, Germany). Probes were prepared by gluing the monodisperse $SiO_2$ beads (d = 25.24 ± 0.75 µm; microParticles GmbH, Berlin, Germany) on tipless silicon nitride cantilevers (MLCT-O10, Bruker, Germany). For each cantilever used, deflection sensitivity was determined by performing a force-distance curve on freshly cleaned glass slide in PBS and actual spring constant was determined by thermal noise method (Hutter and Bechhoefer, 1993) that was ranging from 0.03 to 0.06 N/m. A schematic representation of AFM is shown in Fig. 2a where a bead mounted to a cantilever is indenting a brain tissue immersed in PBS inside a petri dish by applying a precisely controlled load. Schematic representation of the bead of radius R under load $F$ and the indentation depth $\delta$ is shown in Fig. 2b.

Indentation experiments were performed on samples (Fig. 2c) immersed in PBS by mapping the 200 × 200 µm$^2$ area with 8 × 8 pixels as shown in Fig. 2d. All samples from different fresh ovine brains within 6 h post-mortem were investigated. A load of 5 nN was applied on the surface at loading rate of 20 µm/s. After loading, we kept contact for 10 s and change in load ($F$) was recorded at a constant indentation depth ($\delta$).

AFM data was collected in the form of stress relaxation curves that shows how the force decreases with time (for a fixed z-piezo displacement) as a consequence of the internal relaxation process of the WM. Fig. 2e shows a representative example of raw data collected from stress relaxation experiment. Fig. 2e is divided into five regions separated by red dotted lines: (1) approach of the AFM probe to the tissue surface, (2) contact with surface and ramp up to a fixed setpoint, (3) maintenance of constant $\delta$ and the resultant change in $F$, (4) ramp down and (5) retraction of the AFM probe from the tissue surface. All AFM data was initially processed, and quality controlled using JPK data processing software (JPK Instruments AG, Berlin, Germany). All stress-relaxation curves where $\delta$ did not maintain a constant value in time window of 10s were discarded. Then batches of stress-relaxation curves were analysed using MATLAB. In the literature some indentation experiments have considered both ramp and hold sections (section (2) and (3) in Fig. 2e) in order to extract viscoelastic properties (Chen et al., 2020; Qiu et al., 2020, 2018); however, here we deemed the relaxation section (section (3) in Fig. 2e) to be sufficient for the estimation of the time-dependent mechanical response of the tissue. A window of interest was defined for data analysis, window of interest is the region where $\delta$ is maintained at a constant value and the resultant changes in $F$ with respect to time is measured as marked by region (3) in Fig. 2e.

In our experiments, $\delta$ is small compared to the bead radius and the sample thickness; so, at least in first approximation, we can assume a linear elastic and incompressible behaviour at the small scale and, assuming initially nearly isotropic material based on Budday *et al.* (Budday et al., 2017a) findings, we use Hertz model for data analysis. From experimental data, together with the knowledge of the geometry of the system and using Poisson ratio $\sigma = 0.5$, the relaxation modulus $G_R(t)$ (Fig. 2f) was calculated in response to a step indentation depth using Eq. 1:

$$G_R(t) = \frac{3}{16\sqrt{R}\delta_\circ^{3/2}} F(t) \qquad (1)$$

where R is the bead radius and $\delta_\circ$ is step indentation depth.

### Statistical analysis

First, normality distribution of data was checked. A Shapiro-Wilk test rejected hypothesis of normality for all samples. Therefore, statistical significance of non-normal distribution dataset was tested with non-parametric test. Kolmogorov-Smirnov test was used to give an informative statistical decision based on all of the characteristics of data distribution. It is sensitive to not only differences in median but also shape and spread of the distribution while Mann-Whitney is sensitive to the median values only.

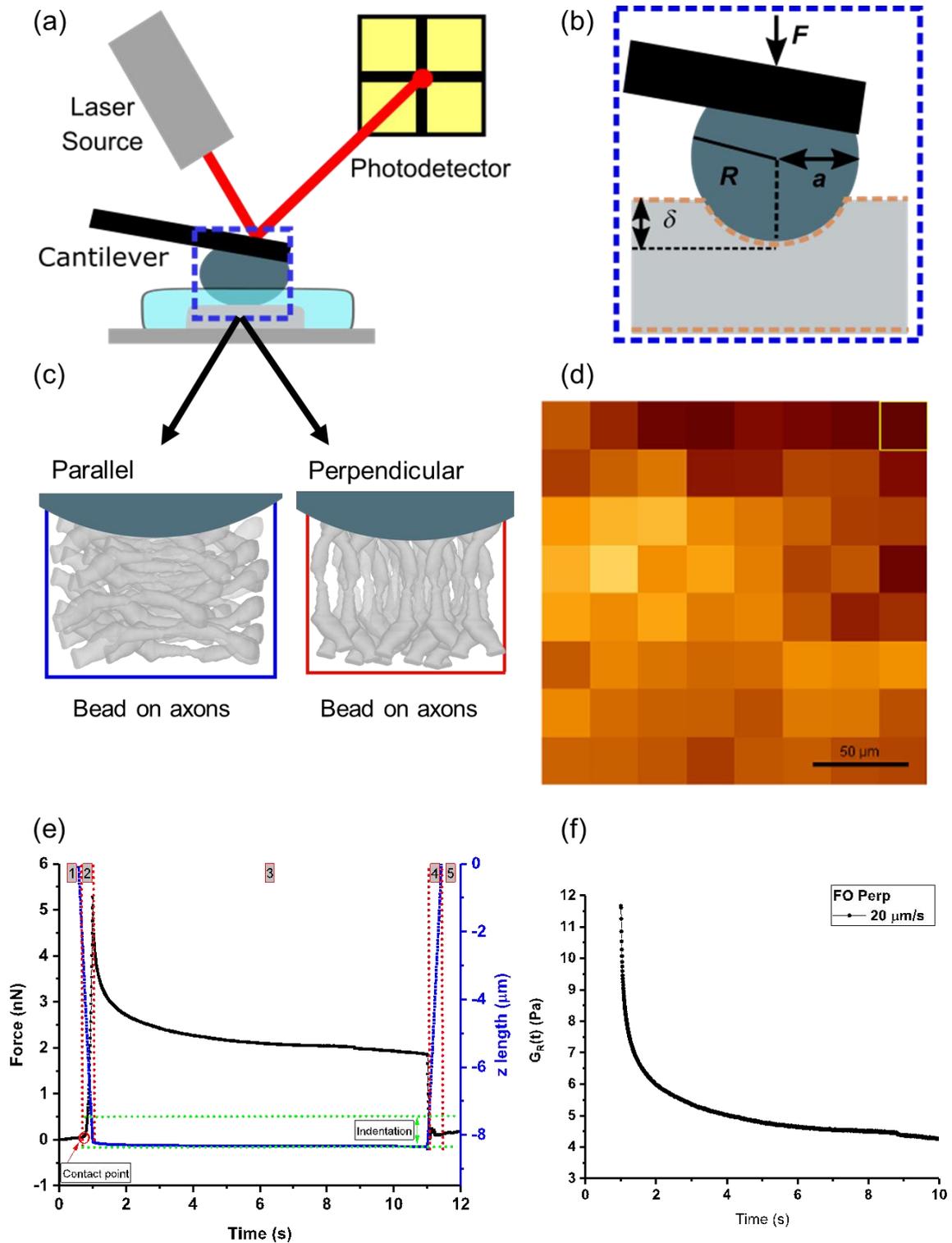

Fig. 2 (a) Schematic representation of an AFM where a bead mounted on cantilever applies load on tissue in petri disch inside PBS and (b) schematic representation of the bead on tissue showing bead geometry and indentation depth δ, as a result of applied load F. (c) Schematic representation of parallel and perpendicular samples, (d) map of a sample with 200 × 200 µm$^2$ area and 8 × 8 pixels where a stress-relaxation curve has been recorded at each pixel. (e) Representative stress-relaxation curves obtained from stress relaxation experiments on brain tissue. Regions separated by red dotted lines and marked by (1), (2), (3), (4) and (5) are the approach of the AFM tip to the sample, ramp up from contact point to the setpoint, control of the constant displacement δ, and corresponding change in F with respect to time, ramp down and retraction of the AFM tip from the sample respectively. Contact point is marked by red circle and indentation depth by green arrow in (e). (f) $G_R(t)$ calculated according to Hertz model.

## 2.3 Micromechanical approach to finite element modelling of WM regions

A reverse-engineering path was followed to find the material parameters of the matrix components *i.e.* axons and ECM, from the experimental relaxation values (section (3) in fig. 2e) treated as representative of the homogenized response of the tissue and its description as a composite material. At the microscale, CC, CR and FO appear as composites characterised by axons as straight reinforcing fibres (Bernardini et al., 2021) and the rest of accessory cells, matrix and intracellular matter acting as the embedding ECM (Javid et al., 2014) resembling a transversely isotropic material, as shown in the focused ion beam electron microscopy (FIB-SEM) image (Bernardini et al., 2021) (Fig. 3a). Each area is characterised by a unique schematic representation of the axons with an elliptical cross-sectional area (***A***), elliptical diameter (***d***) and a volume fraction ($V_f = \frac{Volume_{axons}}{Volume_{total}}$) in an assumed perfect bonding between the axons and ECM (Yousefsani et al., 2018) (Fig. 3b).

From FIB-SEM images of WM, a simplified standard hexagonal pattern, approximates the complex geometry of the axonal network, was chosen to schematically represent the tissue and a cubic 3D RVE was created per each of the areas with its width (***w***), height (***h***) and length (***l***) dependent on the aforementioned area-specific axonal geometry (Fig. 3a-d). The average volume fraction, the modal value of cross-sectional area and the modal value of the diameter of axons listed in Table 1 were taken from our previous study (Bernardini et al., 2021). The commercial finite element analysis software ABAQUS (Version 6.16, Dassault Systèmes Simulia Corp, 2015) was used for the modelling and analysis of the RVE.

**Table 1**

Geometrical parameters of axons, where *A* is the cross-sectional area, *d* is the elliptical diameter, $V_f$ is the volume fraction, *w* is the width, *h* is the height and *l* is the length. (Bernardini et al., 2021).

| Region | A | d | $V_f$ | w = h = l |
|--------|---|---|-------|-----------|
| CC | 0.63 $\mu m^2$ | 1.07 $\mu m$ | 0.45 | 1.67 $\mu m$ |
| CR | 0.50 $\mu m^2$ | 1.00 $\mu m$ | 0.37 | 1.64 $\mu m$ |
| FO | 0.98 $\mu m^2$ | 1.41 $\mu m$ | 0.51 | 1.96 $\mu m$ |

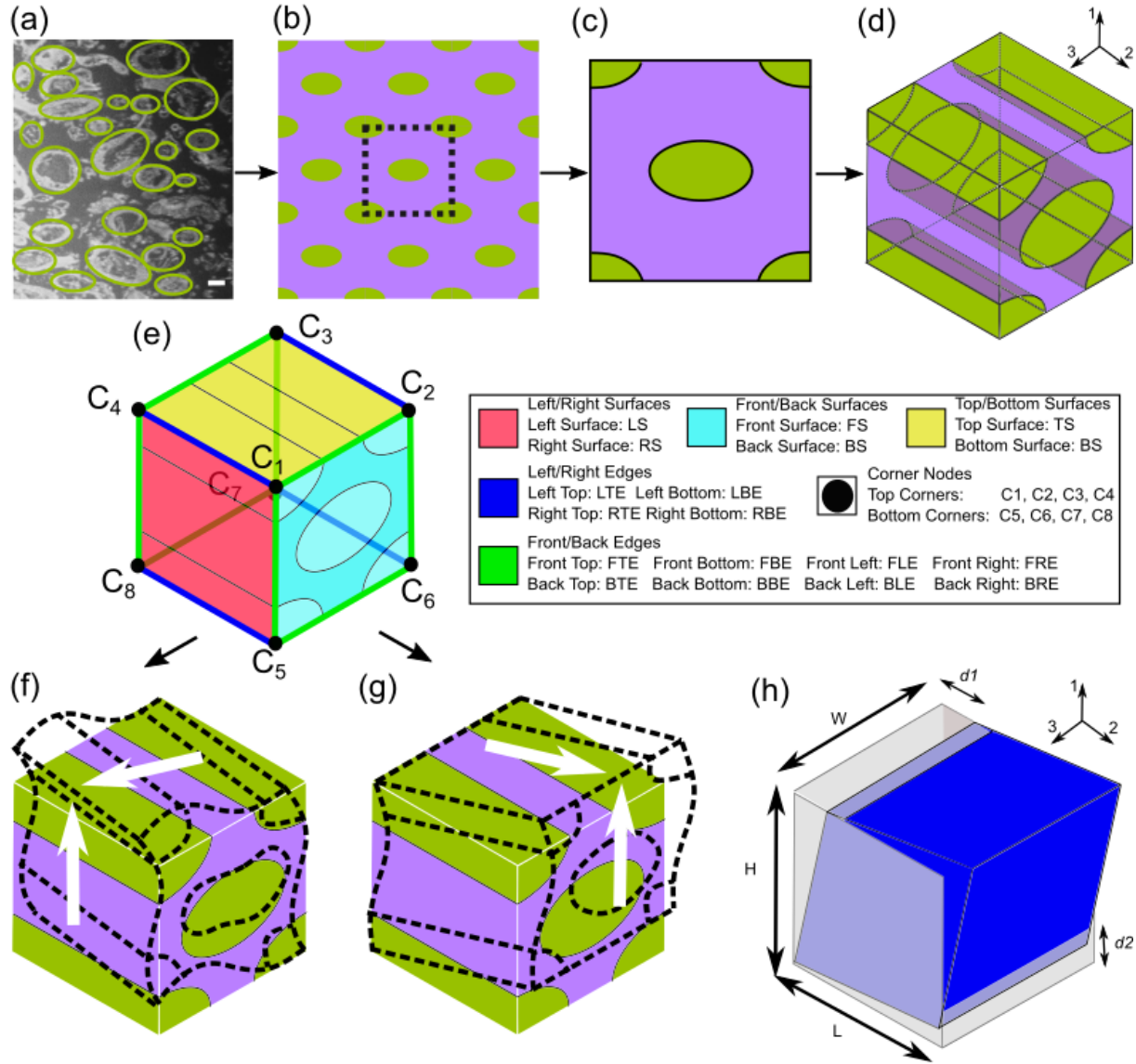

Fig. 3 FIB SEM acquired images (a) [bar scale of 2 µm] are schematised as hexagonal patterned arrangement of the composite (b), from which 2D Representative Constitutive Units (RUCs) (c) are extrapolated with each area characterised by its own axonal geometry. The 3D Representative Volume Element (RVE) follows the $V_f$ measured in each area (d). On each RVE sets of nodes are defined (e) in order to apply PBCs to simulate the $G_{12}$ and $G_{23}$ shear modes (f) (g) by applying the respective displacements (h) on the involved node sets.

### 2.3.1 Calculation of RVE Shear Moduli

In order to simulate the behaviour of the RVE in the continuum, unified Periodic Boundary

Conditions (PBCs) were implemented. The underlying concept of PBCs is that any deformation happening in the RVE studied is the same in any adjacent unit cell. See Supplementary Information for further detail on PBCs and details provided in (Abolfathi et al., 2009; Karami et al., 2009; Omairey et al., 2019). At each time point $\Delta_t = 0.125s$, the shear stress is calculated by dividing the sum of reaction forces of the involved surfaces by their area. Therefore, the shear modulus over time $G(t)$ is then calculated as the ratio between the shear stress and the tensors of the shear strain in each time step. For example, the calculation of $G_{12}$ follows Eq. 2 (see Fig. 3h for visual reference).

$$G_{12} = \frac{\frac{\sum RF_1^{Top}}{L \times W}}{\frac{d_1}{H} + \frac{d_2}{L}} \tag{2}$$

where $\sum RF_1^{Top}$ is sum of the reaction forces of the top surface, $L$, $W$ and $H$ are length, width and height respectively and $d_1$, $d_2$ are displacements.

## 2.3.2 Constitutive Modelling

In order to extrapolate the mechanical properties of the components, the axon and ECM, linear viscoelastic material model (Abolfathi et al., 2009; Javid et al., 2014; Park et al., 2019; Samadi-Dooki et al., 2017) was applied due to its good concordance under the small deformation condition of the experimental method. Therefore, for linear viscoelastic materials the shear modulus over time $G(t)$ can be expressed by a Prony series given by Eq. 3:

$$G(t) = G_0 \left(1 - \sum_{i=1}^{2} g_i \left(1 - e^{-\frac{t}{\tau_i}}\right)\right) \tag{3}$$

Here, $G_0$ is the instantaneous relaxation modulus, $g_i$ is the dimensionless shear modulus *i.e.* characteristic relaxation coefficient and $\tau_i$ is the time relaxation material parameter. Given the observed short relaxation phase and the overall observational time of 10s in our experimental results, a two-term expansion was deemed enough to represent the material ($i = 2$) (Javid et al., 2014).

### 2.3.3 Estimation of axonal and ECM material parameters

With assumption of isotropy of the axons and ECM, their respective material parameters were estimated from the experimental data by solving a minimization problem by Swarm Particle Algorithm (SPA). SPA is being increasingly used in biomechanical research (Ramzanpour et al., 2019) and here, it was used to find the optimal viscoelastic parameters that would minimize the prescribed cost function:

$$Cost = \sqrt{\frac{\sum_{j=1}^{n}\left(1 - \frac{G_j^{est}}{G_j^{exp}}\right)^2}{n}} \qquad (4)$$

where $n = 2049$ is the number of the experimental time points characterized by an acquisition step of $\Delta_t = 0.0049s$, $G_j^{exp}$ is the experimental value of the shear modulus at time point $j$ and $G_j^{est}$ is the analytically estimated value of the shear modulus. The later was calculated by adapting the following assumptions. Because of the small deformations applied experimentally and the very small acquisition step $\Delta_t$, it was assumed that there is no compressive failure associated with the axonal microbuckling (Lo and Chim, 1992). Additionally, in the discreet domain characterized by $\Delta_t$, linear elasticity was assumed. Moreover, given the fluid-filled nature of the cellular components, a near incompressibility described by a Poisson's ratio of $v = 0.499$ was applied. Therefore, one can approximate the shear modulus as being circa a third of the Young's modulus following Eq. 5:

$$E = G * 2(1 + v) \cong 3G \qquad (5)$$

Consequently, analytical estimation of the young's modulus $E$ and therefore of $G$, is achievable by the Rule of Mixture (RoF) (Hyer and Waas, 2000; Jin et al., 2008). Eq. 6 describes the upper and lower estimates of the composite behaviour in the axial and transverse loading, respectively the Voigt and Reuss model (Hyer and Waas, 2000; Reuss, 1929; Sudheer et al., 2015; Voigt, 1889). Namely, these two loadings correspond to our "perpendicular" and "parallel" experimental setup.

$$\begin{cases} G_{perp}^{est} \cong v_{ax} * G_{ax} + (1 - v_{ax}) * G_{ECM} \\ G_{para}^{est} \cong \left(\frac{v_{ax}}{G_{ax}} + \frac{(1 - v_{ax})}{G_{ECM}}\right)^{-1} \end{cases} \quad (6)$$

where $G_{perp}^{est}$ is the analytical estimate of the shear modulus of our "perpendicular" samples and $G_{para}^{est}$ is the one of our "parallel" samples. $v_{ax}$ is the volume fraction of the axons, $G_{ax}$ is the shear modulus of the axons and $G_{mat}$ is the shear modulus of ECM. Therefore, by substituting Eq. 3 in Eq. 6 one can obtain the upper and lower analytical estimates of the Prony expansion of the composite in function of the Prony series of the two components: the axon fibres and the ECM (see Eq. S2 and S3 in Supplementary Information). Therefore, the SPA algorithm, run by a Python script, minimized Eq. 4 by searching and substituting in the equation the optimal Prony parameters of the components. The algorithm would stop when the value of the best position of the cost function would change less than $10^{-08}$ from the global position. The search spaces for Prony parameter of the components, respectively the axons and the ECM, were defined as: $g_{1,2_{ax}} \in [10^{-7}\ 0.99]$, $\tau_{1,2_{ax}} \in [10^{-7}\ 1]$, $g_{1,2_{ECM}} \in [10^{-7}\ 0.99]$, $\tau_{1,2_{ECM}} \in [10^{-7}\ 1]$. For the estimate of the initial shear moduli of the components, the assumption of instantaneous load was adopted, because of the high extension rate of the AFM tip (20 $\mu m/s$). Then, the values experimentally recorded at time $t = 0$ were substituted in Eq. 7 and the system was then solved for CC, CR and FO. Therefore, perpendicular data was used to estimate the Voigt limits, the upper boundaries of the search spaces, while the parallel was used for the Reuss limits, the lower boundaries of the search spaces. The search spaces for the moduli were defined per each area and listed in Table S2 in Supplementary Information:

To mitigate the effect of possible local minima in the search space of the algorithm, 100 estimates of the parameters were performed and from these a weighted average per each area and normalised weights were calculated, see Eq. S4 and S5 in supporting information.

The fitting was performed per each area on each perpendicular and parallel dataset. This has resulted in two sets of parameters for the components: a Voigt and a Reuss estimation. Corresponding analytical force relaxation curves were calculated from their respective Prony series within the experimental period of $t = 10s$ at each acquisition time of $t_j$ with $j = \{0, 0.0049, \ldots, 9.9951, 10\}$ resulting the acquisition step of $\Delta_t = 0.0049s$ to resemble the experimental data. Then, by averaging the perpendicular and parallel analytical data sets (see section S-7 in Supplementary Information), the average analytical datasets corresponding to the Hill estimates, (see Eq. S6) for axonal and ECM data were calculated per each area.

On these time series, a final parameter estimation by fitting of a two term Prony series on $G_{Hill}^{ax}$ and $G_{Hill}^{ECM}$ was performed. The same procedure has been followed: 100 estimates of the Prony parameters were achieved by the SPA through the same python script. This time, the boundaries of the search space for the final parameters have been defined by the first perpendicular (Voigt) and parallel (Reuss) estimates as listed in Table S3 in Supplementary Information. From the 100 estimates, the weighted averages of the parameters were calculated as previously described. These final sets of parameters that describe the Hill estimates of the components, have been used in the FEA simulations as input material parameters of the axonal fibres and the ECM. Finally, the Lin's concordance correlation coefficient (CCC) (Lin, 1992) was calculated by MATLAB to check the agreement between the FEA estimates of $G_{12}$, $G_{23}$ and the experimental values.

## 3 Results

### 3.1 Stress relaxation

Time dependent mechanical behaviour of WM samples was investigated by AFM indentations. Though we investigated the mechanical behaviour at different loading rates but it is known that regional differences are also rate dependent *e.g.* faster loading regimes have highlighted how CR of WM exhibited a much stiffer behaviour when compared to grey matter (GM) and in contrast, slower loading regimes have shown the opposite behaviour (Jin et al., 2013; McCracken et al., 2005). Therefore, we only discuss here the fastest loading rate we adapted in our experiments. For experimental data at lower loading rates, see the Supplementary

Information. Fig. 4a-c shows the stress relaxation behaviour of perpendicular and parallel CC, CR and FO samples recorded at 20 µm/s loading rate. We averaged the relaxation behaviour over all data obtained for a specific type of samples. A difference between the parallel and perpendicular samples from the same regions of WM is observable. Box plots of the peak $G_R(t)$ (at $t = 0$) of the corresponding samples and $G_R(t)$ after 10s as percentage of the peak value are shown in Fig. 4d,e. Shapiro-Wilk test rejected normality for all samples hence non-parametric analysis was carried out for checking statistical significance. The peak value of $G_R(t)$ is higher for perpendicular samples than parallel samples (Fig. 4d), comparing the mean values, that is, for CC by ~15.4%, CR by ~16.8% and FO by ~7.8%. The Kolmogorov-Smirnov test yields this difference is significant irrespective of the WM region *i.e.* CC ($p = 1 \times 10^{-5}$), CR ($p = 5 \times 10^{-17}$) and FO ($p = 2.8 \times 10^{-5}$). The relaxation behaviour of perpendicular and parallel samples was analysed by comparing the percentage of the peak $G_R(t)$ after 10s. The perpendicular samples of CC relax more than parallel samples while this is opposite in FO and apparently no difference in CR (Fig. 4e). This difference between parallel and perpendicular for CC and FO samples is significant ($p = 2 \times 10^{-23}$ and 0.00118 respectively), however; not significant for CR ($p = 0.2$). Overall, CR perpendicular samples appear to be taking longer to fully relax in comparison to other samples (Fig. 4b). This relatively different relaxation behaviour might be due to contribution of imperfect samples as explained in the sample preparation section for CR. Among the different regions of WM, $G_R(t)$ of FO is smaller than CC and CR. The mean values of peak $G_R(t)$ of CR > CC > FO in both perpendicular and parallel samples.

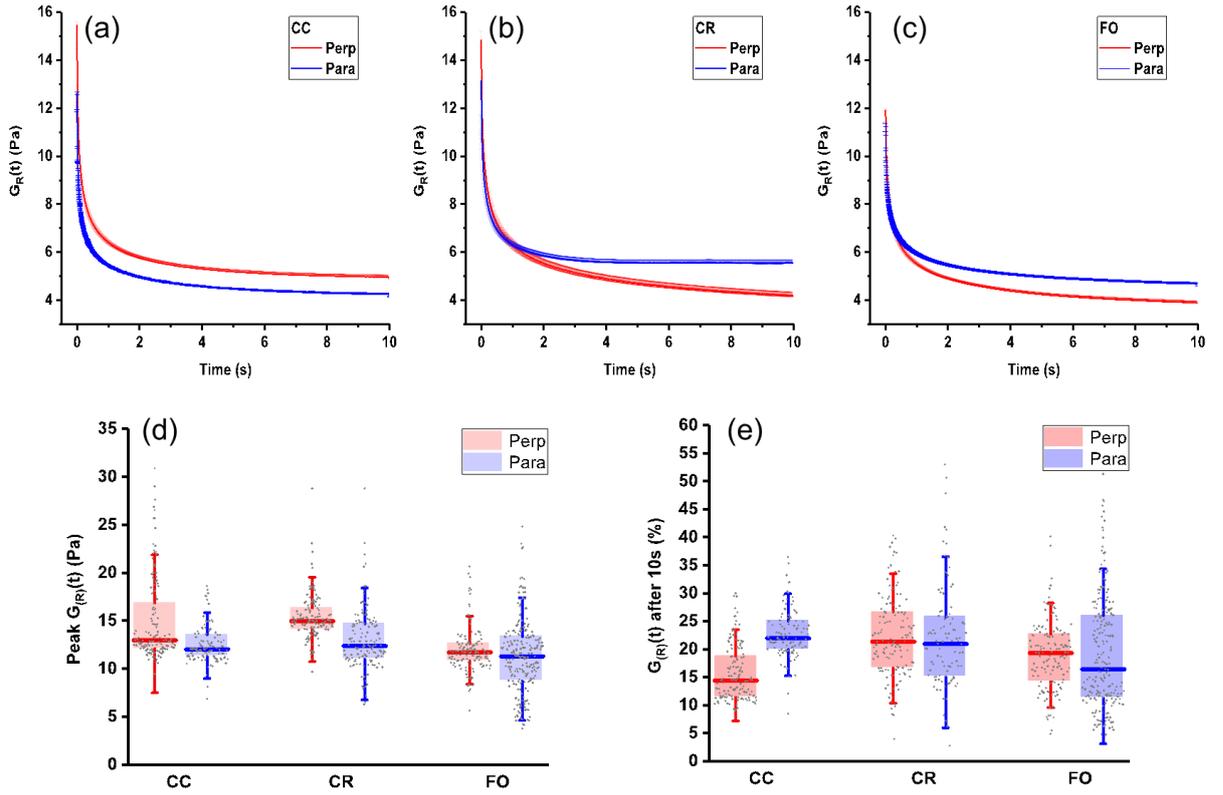

Fig. 4 Time-dependent behaviour of WM tissue: Averaged $G_R(t)$ (solid line) along with SE (shaded region) calculated from experimental data at load = 5nN and loading rate = 20 µm/s of perpendicular and parallel samples of (a) CC, (b), CR and (c) FO. Box plots of $G_R(t)$ at (d) $t = 0$ and (e) at $t = 10s$ as percentage of the peak value. Horizontal line, box, whiskers and grey dots represent the median, Q1-Q3 percentile, 1.5×SD and single data points respectively.

### 3.2 Prony parameters study

The Prony parameters of axons and ECM estimated from experimental data of CC, CR and FO provide an insight in the relative contribution of the components to the tissue homogenised time dependent mechanical behaviour as shown in Table 4. Stiffness of ECM is lower than axons, and Young's modulus E of ECM when compared to axons is ~47%, ~42% and ~25.6% lower in CC, CR and FO respectively. Comparing stiffness of axons in different parts of WM, stiffness of axons in FO is ~32.7% lower than that in CC which is ~1.1% lower than that in CR. However, stiffness of ECM in FO is ~5.4% lower than that in CC which is ~9.7% lower than that in CR. The axons relative contribution to overall stiffness of the tissue appears to be the differentiating factor when elastic behaviour is compared *e.g.* FO with either CC or CR.

Both, axons and ECM relax on short time scale *i.e.* within 2s and overall, for each component, $\tau_1 < \tau_2$. Comparing the relative relaxation behaviour of axons and ECM, in CC $\tau_{1axon} < \tau_{1ECM}$ ~30%, in FO $\tau_{1axon} < \tau_{1ECM}$ ~11%, however; in CR $\tau_{1axon} > \tau_{1ECM}$ ~12%. Also, comparing the relaxation coefficients, in CC $g_{1axons} > g_{1ECM}$ ~22%, in CR $g_{1axons} > g_{1ECM}$ ~0.4% and in FO $g_{1axons} > g_{1ECM}$ ~7%.

**Table 4**

Prony parameters estimated from experimental data.

| Sample | Component | $E_\circ$(Pa) | $g_1$ | $\tau_1$ | $g_2$ | $\tau_2$ |
|---|---|---|---|---|---|---|
| **CC** | Axons | 18.985 | 0.338 | 0.097 | 0.313 | 0.712 |
|  | ECM | 10.037 | 0.263 | 0.126 | 0.348 | 0.714 |
| **CR** | Axons | 19.205 | 0.337 | 0.125 | 0.253 | 0.701 |
|  | ECM | 11.114 | 0.335 | 0.110 | 0.220 | 0.680 |
| **FO** | Axons | 12.773 | 0.210 | 0.145 | 0.338 | 0.664 |
|  | ECM | 9.499 | 0.196 | 0.161 | 0.443 | 0.748 |

### 3.3 Micromechanical modelling

Figure 5 shows analytical estimates of Prony expansion of the composite, $G_{Reuss}$ for parallel samples and $G_{Voigt}$ for perpendicular samples, the FEA estimates $G_{23}$ for parallel samples and $G_{12}$ for perpendicular samples, and the corresponding experimental values $G_{exp}$. The dashed lines represent analytical estimates calculated with Prony parameters at 0.5 times standard deviation from their averaged estimates. All samples show a good agreement among the experimental data, analytical and FEA estimates except the CR perpendicular sample which has small deviation from analytical estimate because of its continued relaxation phase in the experiments not being captured by the two-terms Prony expansion. The ratio between FEA calculated homogenised $E_{11}$ and $G_{12}$, and $E_{33}$ and $G_{23}$ is around 3, with minimal values of 2.6 and maximum values of 3.4 in the FO for example, suggesting the tissue behaves intermittently isotropic. Lin's CCC quantifies agreement between experimental values and FEA estimates, i.e. between $G_{exp}$ and $G_{23}$ for parallel samples and between $G_{exp}$ and $G_{12}$ for perpendicular samples. Please note, we adapted the assumption that our composite is transversely isotropic i.e. in case of perpendicular samples, $G_{12} = G_{13}$ (Koser et al., 2015). For CC, CR and FO samples, Lin's CCC is 0.8205 [0.7620, 0.8658], 0.8940 [0.8562, 0.9223] and 0.8893 [0.8448, 0.9216] for parallel samples and 0.9635 [0.9456, 0.9755], 0.9272 [0.8965, 0.9491] and 0.9302 [0.8981, 0.9525] (mean, [95% confidence interval]) for perpendicular samples respectively.

The averaged relaxation behaviour of perpendicular and parallel samples, represented as Hill estimates, of the analytical, FEA and experimental data are shown in Fig. 6. The analytical and FEA estimates are in good agreement with the experimental data of samples from CC, CR and FO.

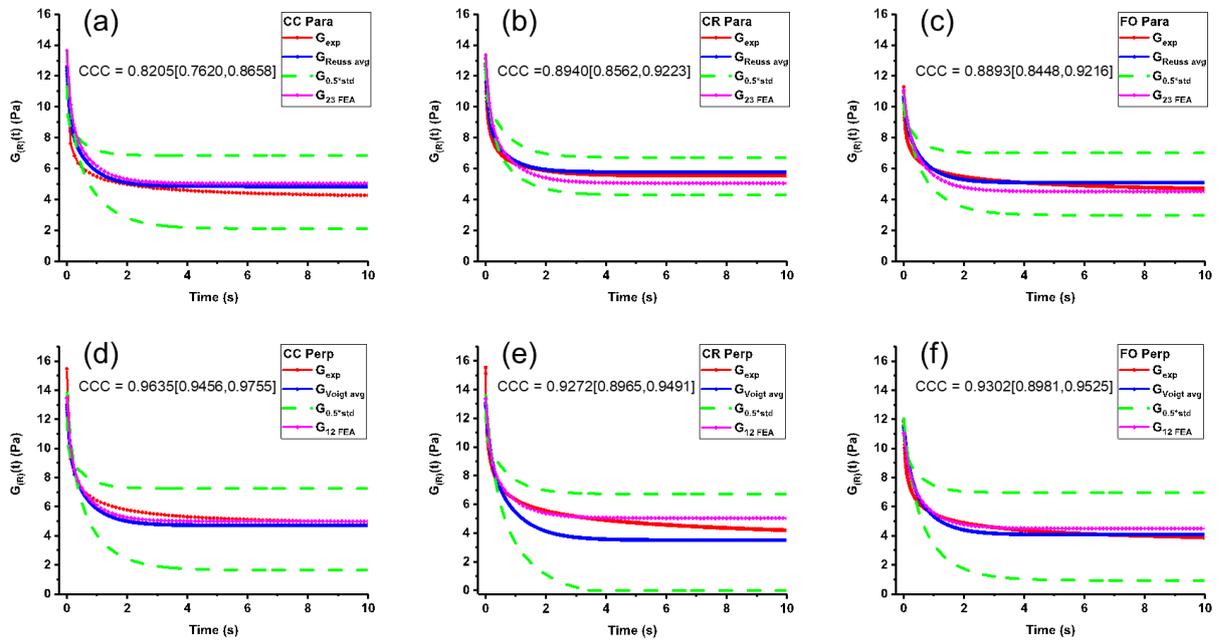

Fig. 5 Analytical and FEA estimates and the experimental values of stress relaxation behaviour of (a, b, c) parallel and (d, e, f) perpendicular samples of CC, CR and FO respectively. CCC represent the agreement between experimental data and FEA estimates.

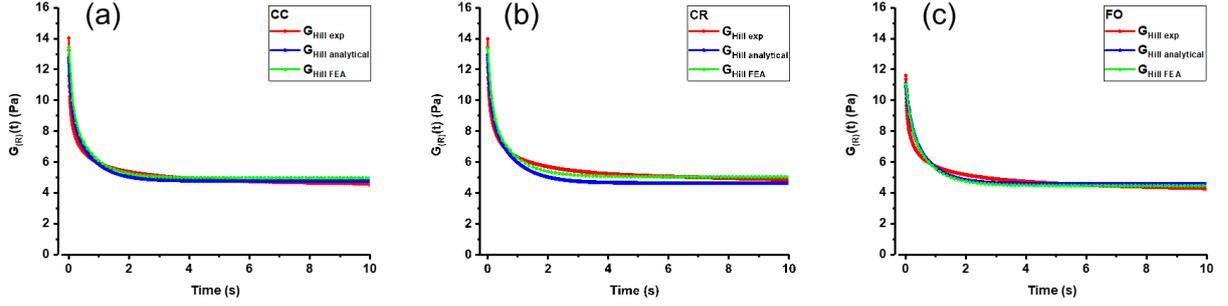

Fig. 6 Hill estimates of the analytical, FEA and experimental data of (a) CC, (b) CR and (c) FO showing a good agreement.

# 4 Discussion

We systematically investigate the time-dependent mechanical behaviour of three regions of ovine WM tissue *i.e.* CC, CR and FO while considering the directionality of axons in those specific regions. We observed a difference in the mechanical behaviour due to orientation of axons from the same regions of WM. When assessed at $t = 0$, the perpendicular samples are stiffer than parallel samples however this trend does not correlate with their relaxation behaviour when assessed at $t = 10$s as percentage of the peak $G_R(t)$.

In WM tissue, the effect of directionality of axons on its mechanical behaviour has been reported in literature; however, limited understanding has been developed because of the absence of definitive and uncontroversial results. Several factors including the type of measurements, sample properties, and physiological conditions potentially contribute to these differences reported in literature.

Budday *et al.* (Budday et al., 2017a) investigated cubic samples of ~5 mm side length of human brain tissue. Considering directionality of axons in CC of WM tissue, they applied compression and tension in two direction (D1, D2) and shear in three orthogonal directions (D1, D2, D3), see Fig. 7. Though their data hints a marginally softer average mechanical response along the axons direction in compression (0.42 kPa and 0.5 kPa along D1 and D2 respectively) and stiffer along the axons in tension (0.35 kPa and 0.32 kPa along D1 and D2 respectively), the mean tissue response did not show significant directional dependency. However, it has been shown that WM is normally in tension (Xu et al., 2010, 2009). This important finding may play a role in assessing mechanical behaviour and should be considered in relevant experiments. Also, in shear mode, the results from Budday and co-workers did not show a significant directionality effect in averaged shear moduli *i.e.* 0.37 kPa, 0.29 kPa and 0.33 kPa along D1, D2 and D3 respectively. However; the study by Prange *et al.* (Prange and Margulies, 2002) on porcine WM prismatic samples (10×5×1 mm) reports a higher shear response in orthogonal to the axons direction (D2) than in parallel (D1) in CC samples ($\mu_\circ$ = 232.2 Pa and 131.4 Pa respectively), and opposite findings in CR samples ($\mu_\circ$ = 292.6 Pa and 211.4 Pa along D1 and D2 respectively).

In contrast, treating porcine and lamb WM as transversely isotropic material, Feng *et al*. (Feng et al., 2017a, 2013) have reported higher stiffness of CC samples when axons orientation was parallel to the shear direction (D2) than that when axons were oriented perpendicular (D3). They have also shown with an asymmetric indenter that stiffness is dependent on the relative orientation of axons and long side of indenter. In the same plane (like D2 in Fig. 7(a)), stiffness of CC was higher when long side of a rectangular indenter was perpendicular to axon's direction than that in parallel *i.e.* indentation stiffness 170.5 mN/mm and 72.9 mN/mm respectively. For a full comparison of experimental data in literature on the brain mechanical properties, see Table S1 in Supplementary Information.

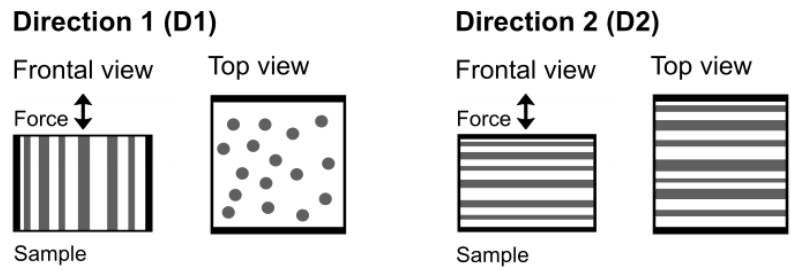

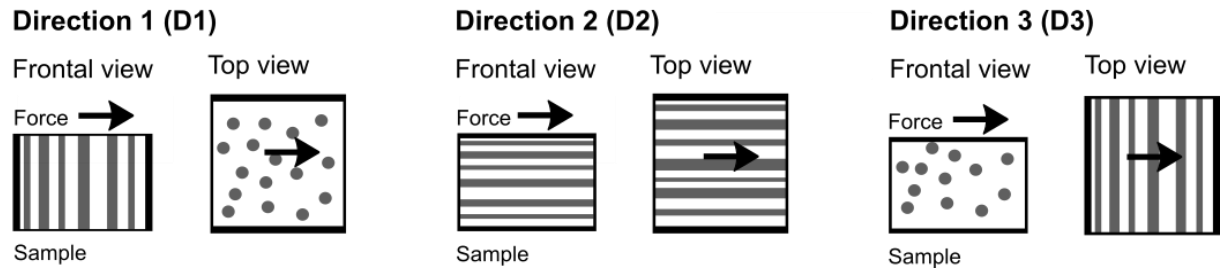

Fig. 7 Schematic representation providing an overview of force relative to axons orientation in mechanical testing. It shows two possible compression and tension modes and three possible shear modes for a WM sample with uniaxially oriented axons. In our case, D1 corresponds to perpendicular samples and D2 to parallel samples.

These studies have shown a clear appreciation for the effect of WM anisotropy on its mechanical behaviour; however, they are yet inconclusive in terms of a quantitative assessment of to the contribution of the constitutive elements of the complex tissue to its specific mechanical behaviour as a function of axons orientation. Please note their macroscopic samples, macroscale scale measurements and dependence on the indenter shape. This makes it difficult to capture microstructural heterogeneities present in different regions of WM tissue that potentially contribute to contradicting results (Budday et al., 2017a). Furthermore, macroscale indentation experiments on large samples as confined tests keep fluid trapped (Budday et al., 2015), and increased stiffness of the tissue due to fluid phase contribution cannot be ruled out (Budday et al., 2017b).

The AFM indentations performed in this study are at the length scale relevant to tissue components and its micron sized tip should prevent fluid entrapment and minimise interactions with the solid phase (Budday et al., 2019); this in theory decouple the response of solid and fluid phases and enables us to focus on the response of the tissue without having to consider poroelastic nonlinear effects. Also the directional axons in WM provide preferential pathways for fluid flow (Lonser et al., 2005). Based on this, AFM indentations on WM in the small strain limits can be treated as unconfined tests and, like studies characterised by slow displacement rates, show softer mechanical response.

The mechanical stiffness results obtained here agree with similar AFM-based tests that also showed a very soft mechanical response, in the order of Pa, for WM tissue (Koser et al., 2015; Moeendarbary et al., 2017). In a similar compressive AFM study in low strain limits on spinal cord, composed of even more organised WM, Koser *et al.* (Koser et al., 2015) showed lower stiffness when axons were oriented along the force direction (D1) than that perpendicular to force direction (D2). The observed directionality effect of axons in spinal cord WM is in contrast to our results of brain WM, where axons orientation along the force direction yields higher peak value of $G_R(t)$; however, these findings agree that, regardless of its area of origin, WM appears to behave at the microscale as an ultrasoft transversely isotropic material. Furthermore, it is not known how much spinal cord WM resembles brain WM in terms of fundamental tissue components that affect mechanical behaviour such as axons geometry and its relative axons volume fraction. Although they are both WM tissues, a specific cytoarchitectural comparison is yet to be performed and specific cytoarchitectural differences are currently unknown. Overall, spinal cord has a more organised and unidirectional nature with extremely long axons (Kakulas et al., 1998) due to its macroscopic arrangement so it might be speculated to have a different volume fraction of WM causing the difference in behaviour. Therefore, we believe these apparent discrepancies in mechanical behaviour may be explained by further studying such highly complex geometrical parameters and understanding the relative contribution of individual tissue components.

We used reverse engineering approach to extract Prony parameters from our experimental data which reveal the individual mechanical behaviour of WM tissue components *i.e.* axons and ECM. Knowing the mechanical behaviour of axons and ECM is crucial for the development of realistic mechanical models (Elkin et al., 2011a, 2011b; Finan et al., 2012; Lee et al., 2014). Attempts have been made by other researchers to use different approaches, such as magnetic resonance elastography (MRE), for the in vivo non-invasive estimation of material parameters, *e.g.* the Holzapfel–Gasser–Ogden (HGO) parameters (Hou et al., 2020), considering the biological tissue as a soft porous fibrous material. However, resolution limits make it difficult to pinpoint and characterise small and irregularly shaped areas of the brain such as the CC or the FO, which are both areas of interest in this study (Papazoglou et al., 2012). Therefore, although promising and with a great potential, MRE and other in vivo tools currently cannot substitute the fundamental complementary role played by *ex vivo* studies (Bilston, 2018).

Previously, mechanical properties of axons and ECM have been investigated experimentally, however; most of these studies were performed on components taken out from their natural environment or grown in-vitro. For example, Ouyang *et al.* (Ouyang et al., 2013) denatured protein components and investigated the mechanical properties of single intact axon, axon with disrupted microtubules and disrupted microfilament. They probed a single axon (radius ~1 µm) on a surface using AFM (deformation ~1 µm), which is a very different environment than an axon embedded in ECM.Similarly, the mechanical environment affecting behaviour of cells has been well documented (Bertalan et al., 2020; Blumenthal et al., 2014; Iwashita et al., 2014; Jaalouk and Lammerding, 2009; Schlüßler et al., 2018; Wozniak and Chen, 2009). Therefore, to understand the WM mechanics at the continuum level and overcome the discrepancies found in literature *e.g.* in ECM mechanics (Young et al., 2016), it is important to extract tissue components mechanical parameters when they are in their natural environment. Our extrapolated axons and ECM parameters obtained while the tissue components are in their own natural environment overcome this issue and provide a bottom-up understanding of how the tissue underlying cytoarchitecture plays a role in the homogenised mechanical behaviour of this complex tissue.

Furthermore, to develop micromechanical model of brain WM, we used the components parameters and calculated the analytical and FEA estimates, which are in good agreement with our experimental data. The estimates of the axonal and ECM material parameters, can then be used to create region specific models of any WM areas, once knowing the respective $V_f$ and geometrical properties of the axons, and predict the homogenized mechanical properties of such a complex tissue. It should also be emphasised that the data obtained here and the micromechanical characterisation performed in this paper may help disentangling the complex interplay between porous and viscous effects observed in brain tissue (Comellas et al., 2020).

## 5 Conclusions

The localised microscale mechanical behaviour of brain WM, specifically when accounted for tissue anisotropy, has so far not been systematically studied. In this work, we investigated the effect of local microstructural features on the time-dependent mechanical behaviour of three regions of brain WM, *i.e.* CC, CR and FO in a systematic manner using AFM indentation. We accounted for the directionality of axons in these three regions of WM if there is a dependence of mechanical behaviour on anisotropy of WM tissue.

Our results demonstrate that the tissue stiffness is higher when axons are oriented along the applied load direction than when the load is perpendicular to axons for all three regions of WM. Furthermore, the relaxation behaviour is also different when the axons orientation is different in CC and FO regions of WM. Importantly; we have extracted the Prony parameters of axons and ECM of three separate regions, *i.e.* CC, CR and FO, from their homogenised microscale experimental data. We have used this data to develop micromechanical model of brain WM and have calculated the analytical and FEA estimates of CC, CR and FO.

Our experimental investigations combined with micromechanical modelling approach provide much needed information about the localised mechanical behaviour of WM as a function of the orientation of the WM fibre bundles. This provides evidence of the need to include WM tissue anisotropy as one of the key parameters for the optimisation of smart surgical systems such as CED and to develop realistic models and simulate mechanical behaviour of brain WM to understand brain development and diseases.

## Acknowledgements

This work was supported by EDEN2020, a project funded by the European Union's H2020 Research and Innovation Programme under Grant agreement No. 688279. DD also acknowledges support received from the Engineering and Physical Sciences Research Council (EPSRC) through his Established Career Fellowship EP/N025954/1.

# Supplementary Information

# Microscale characterisation of the time-dependent mechanical behaviour of brain white matter

Asad Jamal [a,*,#], Andrea Bernardini [a,#] and Daniele Dini [a]

[a] *Department of Mechanical Engineering, Imperial College London, London, SW7 2AZ, UK*

[*]Correspondence e-mail: a.jamal@imperial.ac.uk

[#]These authors contributed equally to the work presented in this article

**S-1. Summary of mechanical characterisation of brain tissue**

**Table S1.** Experimental studies on the mechanical characterisation of brain tissue in literature

| Species | Tissue | Sample area | Sample size | Time post-mortem | Experimental technique | Mechanical properties | Study |
|---|---|---|---|---|---|---|---|
| Porcine | WM | CC | H = 3.3 mm<br>D = 14 mm | 4h | Macro indentation (indenter: 2×20 mm) | Indentation stiffness:<br>$E_{para}$ = 72.9, $E_{perp}$ = 170.5 mN/mm | (Feng et al., 2017) |
| Porcine | WM | CC | | in-vivo<br>in-situ | Elastography | Storage and loss moduli<br>$G'$ = 1.22 to 4.57 kPa<br>$G''$ = 0.43 to 2.26 kPa | (Weickenmeier et al., 2018) |
| | | | H = 5 mm | ex-vivo | Hysteron TI 950 Triboindenter (Flat punch = 1.5 mm) | Elastic shear modulus<br>G = 0.28 to 0.38 kPa | |
| Human | WM | CC, CR | W = H = L = 5 mm | 24h | Macro shear, tension, compression | Elastic shear modulus<br>G = 0.4 to 1.4kPa | (Budday et al., 2017a) |
| | GM | Cortex, Basal ganglia | | | | | |
| Human | WM | CR | W = H = L = 5 mm | 48h | Macro shear, tension, compression | Zener model, Maxwell representation<br>$\mu_\infty = 0.3\ kPa, \mu_1 = 0.9\ kPa, \tau_1 = 14.9\ s$ | (Budday et al., 2017b) |
| | GM | Cortex | | | | Zener model, Maxwell representation<br>$\mu_\infty = 0.7\ kPa, \mu_1 = 2.0\ kPa, \tau_1 = 9.7s$ | |
| Mouse | WM | Spinal cord | H = 500 μm | 6 - 7h | AFM<br>Tip diameter = ~37 μm | Elastic modulus:<br>E ~ 70 Pa | (Koser et al., 2015) |
| | GM | | | | | Elastic modulus:<br>E ~130 Pa | |
| Bovine | WM | Various cerebrum locations | H = 5 mm | 6h | Hysteron TI 950 TriboIndentor (Flat punch = 1.5 mm) | Elastic modulus:<br>E = 1.895 ± 0.592 kPa | (Budday et al., 2015) |
| | GM | | | | | Elastic modulus:<br>E = 1.389 ± 0.289 kPa | |
| Rat | WM + GM | Frontal and parietal lobes | H = D = 10 mm | 2h | Macros tension, compression | Elastic modulus:<br>E ~15 to 25 kPa | (Karimi and |
| | | | | | | | (Navidbakhsh, 2014) |
| Ovine | WM | CC | H = 2.8 mm, D = 15.6 mm | 5h | Macro indentation (rectangular indenter 19.1×1.6 mm) | Indentation stiffness: $K_{para}$ ~25 mN/mm $K_{perp}$ ~65 mN/mm | (Feng et al., 2013) |
| | GM | Temporal lobe proximity of cerebellum | | | | Indentation stiffness: K = 25 -80 mN/mm | |
| Porcine | WM + GM | Various cerebrum locations | W = H = 19 mm L = 4 mm | 8h | Macro shear | Prony expansion Initial shear modulus μ = 4942.0 Pa Prony parameters: $g_1$= 0.520, $g_2$ =0.3057 $\tau_1$ = 0.0264 s, and $\tau_2$ = 0.011 s | (Rashid et al., 2013) |
| Human | WM | CC and CR | W = H = 14 mm L = 5 mm | | Macro tension, compression, shear, directionality | No directionality effect in tension and compression Compression stress: CC ~10-20 kPa Compression stress: CR ~16-27 kPa Shear stress along the fibres > perpendicular fibres | (Jin et al., 2013) |
| | GM | Cortex, Thalamus | | | | | |
| Bovine | WM | CC | H = 4 mm, D = 20.5 mm | | Macro compression | Elastic modulus: E=350 Pa | (Cheng and Bilston, 2007) |
| Human | WM | Various cerebrum locations | | in-vivo | MRE | Shear stiffness: 13.6 kPa | (Kruse et al., 2008) |
| | GM | | | | | Shear stiffness: 5.22 kPa | |
| Porcine | WM + GM | Thalamus | H = 1.5 to 3.5 mm D = 10 to 13 mm | 2.5 -10h | Macro shear - Rheometry | Shear modulus G = 200 – 300 Pa | (Garo et al., 2007) |
| Porcine | WM | CR | H = 0.15 to 2.25 mm D = 10 to 20 mm | 1-2 days | Rheometry | Shear modulus: $G = $ ~24.4 $to$ 1.0 $kPa$ between $10^{-5}$ s and 270 s | (Nicolle et al., 2005) |
| Porcine | WM | CR, CC | W = 10 mm H = 1 mm L = 5 mm | 5h | Macro shear, compression directionality | CR stiffer along fibres direction CC stiffer perpendicular to fibres direction $\mu_{peak}$ ~130 to 230 Pa | (Prange and Margulies, 2002) |
| | GM | Thalamus | | | | $\mu_{peak}$ ~250 to 270 Pa | |
| Human | GM | Temporal cortex | | ex-vivo 3h | Macro shear, compression | $\mu_{peak}$ ~295 Pa | |

*: CC = corpus callosum, CR = corona radiata, WM = white matter, GM = grey matter, W = width, D = diameter, H = height.

**S-2. Periodic boundary conditions for RVE**

The Periodic Boundary Conditions (PBC) defined in the ABAQUS plugin tool, node-to-node linear constraints equation to the nodal degrees of freedom were applied together with boundary conditions to prevent rigid body motion (Abolfathi et al., 2009; Karami et al., 2009; Omairey et al., 2019). For example, in simulating shear $G_{12}$ the Degrees of Freedom (DoF) are linked between each pair of nodes following the relationship in Eq. S1 where $u_1\ u_2\ u_3$ represent the displacement over the X, Y and Z direction (see Fig. 3e for right-left, top-bottom and front-back positioning) and $\Delta_{1,2}$ represents the displacements applied on the respective direction. The applied displacements correspond to a final strain of 0.2 when divided by the length of the RVE along the affected axis.

$$\begin{cases} u_1^{Front,Left} - u_1^{Back,Right} = 0 \\ u_2^{Front} - u_2^{Back} = \Delta_2 \\ u_1^{Top} - u_1^{Bottom} = \Delta_1 \\ u_2^{Top,Left} - u_2^{Bottom,Right} = 0 \\ u_2^{Front,Top,Left} - u_2^{Back.Bottom,Right} = 0 \end{cases} \quad (S1)$$

## S-3. Estimation of axonal and ECM material parameters

Substituting Eq.4 into Eq.7 give Eq. S2 and S3. They represent the upper and lower analytical estimate of the Prony expansion of the composite in function of the Prony series of the two components: the axon fibres (subscript ax) and the ECM (subscript ECM).

$$G_{perp}^{est} \cong v_{ax} * G_{0_{ax}} \left( (1 - g_{1_{ax}} - g_{2_{ax}}) + g_{1_{ax}} * e^{-t/\tau_{1_{ax}}} + g_{2_{ax}} * e^{-t/\tau_{2_{ax}}} \right)$$

$$+ (1 - v_{ax}) * G_{0_{ECM}} \left( (1 - g_{1_{ECM}} - g_{2_{ECM}}) + g_{1_{ECM}} * e^{-t/\tau_{1_{ECM}}} + g_{2_{ECM}} \right. \quad (S2)$$

$$\left. * e^{-t/\tau_{2_{ECM}}} \right)$$

$$G_{para}^{est} \cong \left( \frac{v_{ax}}{G_{0_{ax}} \left( (1 - g_{1_{ax}} - g_{2_{ax}}) + g_{1_{ax}} * e^{-t/\tau_{1_{ax}}} + g_{2_{ax}} * e^{-t/\tau_{2_{ax}}} \right)} \right.$$

$$\left. + \frac{v_{ECM}}{G_{0_{ECM}} \left( (1 - g_{1_{ECM}} - g_{2_{ECM}}) + g_{1_{ECM}} * e^{-t/\tau_{1_{ECM}}} + g_{2_{ECM}} * e^{-t/\tau_{2_{ECM}}} \right)} \right)^{-1} \quad (S3)$$

## S-4. Search spaces for moduli of axons and ECM

**Table S2**

Search spaces for moduli of axons and ECM from different regions of WM.

| Sample | Component | $G_0$ (Pa) |
|---|---|---|
| CC | Axons | [8.42 32.14] |
|  | ECM | [9.19 21.25] |

| | | | |
|---|---|---|---|
| CR | Axons | [9.08 21.90] | |
| | ECM | [10.70 18.24] | |
| FO | Axons | [9.24 14.60] | |
| | ECM | [9.15 14.72] | |

## S-5. Weighted average and normalised weight calculations

Weighted average was calculated using Eq. S4

$$\bar{x} = \sum_{i=1}^{100} x_i \widehat{w_i} \tag{S4}$$

where $\bar{x}$ is the weighted average of each parameter, $x_i$ is the $i^{\text{th}}$ estimate of the parameter and $\widehat{w_i}$ are the normalized weights calculated according to Eq. S5

$$\widehat{w_i} = \frac{(Cost_i)^{-1}}{\sum_{i=1}^{100}(Cost_i)^{-1}} \tag{S5}$$

where, $Cost_i$ is the value returned by the cost function at the $i^{\text{th}}$ estimate of the parameter set.

## S-6. Boundaries of the search spaces for final Prony parameters

**Table S3**
Boundaries of the search spaces for finding the final Prony parameters.

| Sample | Component | $G_0$ (Pa) | $g_1$ | $\tau_1$ | $g_2$ | $\tau_2$ |
|---|---|---|---|---|---|---|
| CC | Axons | [18.53 19.46] | [0.32 0.35] | [0.09 0.10] | [0.29 0.33] | [0.64 0.78] |
| | ECM | [9.90 10.20] | [0.26 0.27] | [0.09 0.16] | [0.31 0.37] | [0.70 0.72] |
| CR | Axons | [16.80 18.78] | [0.30 0.35] | [0.09 0.16] | [0.26 0.28] | [0.60 0.73] |
| | ECM | [11.10 11.23] | [0.27 0.29] | [0.10 0.15] | [0.23 0.50] | [0.72 0.87] |
| FO | Axons | [12.02 13.53] | [0.17 0.22] | [0.12 0.15] | [0.33 0.35] | [0.57 0.71] |
| | ECM | [9.49 9.53] | [0.14 0.25] | [0.11 0.19] | [0.38 0.49] | [0.72 0.76] |

## S-7. Hill estimates

The average analytical data sets were calculated by averaging perpendicular and parallel analytical data sets using Eq. S6

$$\begin{cases} G_{Hill}^{ax}(t_j) = \dfrac{G_{perp}^{ax}(t_j) + G_{para}^{ax}(t_j)}{2} \\ G_{Hill}^{ECM}(t_j) = \dfrac{G_{perp}^{ECM}(t_j) + G_{para}^{ECM}(t_j)}{2} \end{cases} \quad (S6)$$

where

$$\begin{cases} G_{perp}^{ax}(t_j) = G_{0\,perp}^{ax}\left(1 - \sum_{i=1}^{2} g_{i\,perp}^{ax}\left(1 - e^{-\frac{t_j}{\tau_{i\,perp}^{ax}}}\right)\right) \\ G_{para}^{ax}(t_j) = G_{0\,para}^{ax}\left(1 - \sum_{i=1}^{2} g_{i\,para}^{ax}\left(1 - e^{-\frac{t_j}{\tau_{i\,para}^{ax}}}\right)\right) \end{cases} \quad (S7)$$

$$\begin{cases} G_{perp}^{ECM}(t_j) = G_{0\,perp}^{ECM}\left(1 - \sum_{i=1}^{2} g_{i\,perp}^{ECM}\left(1 - e^{-\frac{t_j}{\tau_{i\,perp}^{ECM}}}\right)\right) \\ G_{para}^{ECM}(t_j) = G_{0\,para}^{ECM}\left(1 - \sum_{i=1}^{2} g_{i\,para}^{ECM}\left(1 - e^{-\frac{t_j}{\tau_{i\,para}^{ECM}}}\right)\right) \end{cases} \quad (S8)$$

## S-8. Stress-relaxation behaviour at different loading rates

A general trend of lower $G_R(t)$ at higher loading rate can be seen for all samples irrespective of the directionality of the axons and areas. At lower loading rates, $G_R(t)$ varies between different areas and relatively more scattered data, such as FO parallel sample at 10 µm/s in Fig. S1.

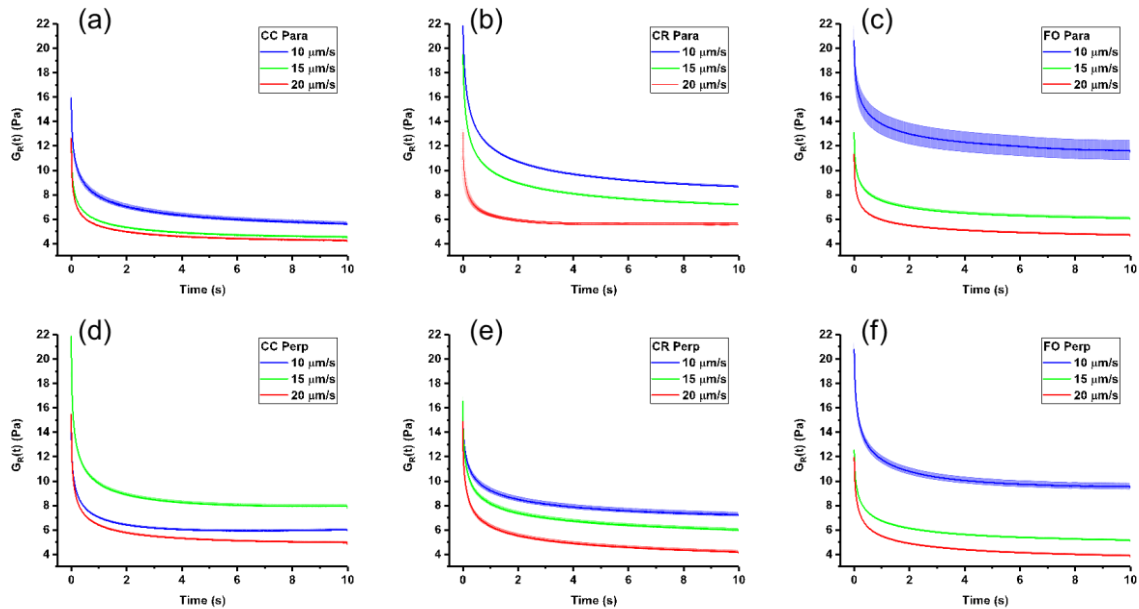

**Fig. S1.** Representative averaged $G_R(t)$ (solid line) along with SE (shaded region) calculated from experimental data for (a-c) parallel and (d-f) perpendicular samples of CC, CR and FO at loading rates of 10, 15 and 20 µm/s.